\input{aipcheck.tex}

\newcommand\sps{\space\space\space\space}
\def\selectedoptions{final}

\documentclass[
   \selectedoptions
  ]
  {aipproc}

\def\selectedlayoutstyle{6x9}
\layoutstyle\selectedlayoutstyle

\SetInternalRegister\hbadness{8000} 

\newcommand\doingARLO[2][]{%
  \ifx\mmref\undefined #1\else #2\fi
}
  
\begin{document}

\title{Performance of the PrimEx Electromagnetic Calorimeter}


\author{M.Kubantsev}{
  address={Institute of Theoretical and Experimental Physics, 
Moscow, Russia/Northwestern University, Evanston, Il 60208, USA},
}

\iftrue
\author{I. Larin}{
  address={Institute of Theoretical and Experimental Physics, 
Moscow, Russia},
}

\author{A. Gasparian}{
  address={Physics Department NC A\&T State University Greensboro, NC 27411, USA},
}


\kern+1cm
\
\kern-1cm
\vspace{+4.5cm}
\begin{center}

for PrimEx Collaboration
\end{center}
\vspace{-4.25cm}

\copyrightyear  {2006}

\date{2006/07/05}

\begin{abstract}
\normalsize
We report the design and performance of the hybrid 
electromagnetic calorimeter consisting of 1152 $PbWO_4$ crystals 
and 576 lead glass blocks for the PrimEx experiment 
at the Jefferson Laboratory. The detector was built for 
high precision measurement of the neutral pion lifetime 
via the Primakoff effect. Calorimeter installation and 
commissioning was completed with the first physics run 
in fall of 2004. 
We present the energy and position resolution of the calorimeter.
Obtained $\pi^0$ mass resolution of $1.3 \mathrm{MeV/c^2}$ and 
its production angle resolution of $0.34 \mathrm{mrad}$ 
demonstrate the ability of the experiment to extract the
$\pi^0$ lifetime on one percent level.     
\end{abstract}
\maketitle
\section{Introduction}
  Measurement of the partial width of the 
  decay $\pi^{0} \to \gamma\gamma$ provides an unique test of low energy 
  QCD in confinement scale  regime. The $\pi^0$ lifetime is arguably the 
  most precise theoretical calculation in low energy QCD,
  which in the leading order depends on fundamental parameters only.
  The contribution of the next-to-leading order does not exceed 4\%
  of the leading-order prediction 
  \cite{Jose_Goity}.
   The experiment PrimEx
  \cite{prim_prop}, \cite{prim_cdr}  was designed to 
  make a high precision (at one percent level) measurement of the $\pi^0$ decay width.
  The tagged photon beam at Hall B of the Jefferson Laboratory \cite{sober}
  was used for 
  forward production of neutral pions in the Coulomb field of the target 
  nucleus (Primakoff effect). The two photons from the pion decay are 
  detected in the Hybrid Calorimeter (HYCAL). The HYCAL was designed to 
  provide high efficiency detection of the photons with high resolution 
  in energy and space. Here we describe the physics requirements 
  for the calorimeter, the design and construction of the setup, and its 
  calibration and performance during the physics data taking. 
  We summarize calorimeter characteristics in the conclusion.
\section{Physic Requirements}
  Precision measurement of an absolute value of the Primakoff $\pi^{0}$ production
  cross section requires a photon detector with high energy and space
  resolutions. Yet another condition is accurate 
  monitoring of the energy and intensity of the tagged photon beam. 
  The Primakoff production cross section 
  is peaked at  $\sim0.3 \mathrm{mrad}$ for a nucleon target at energies of 
  this experiment ($E_{\gamma}$ = 4.9--$5.5 \mathrm{GeV}$). Therefore, high angular 
  resolution is required for clear separation of this mechanism  
  from the other processes at forward angles. Likewise, accurate reconstruction 
  of the two-photon invariant mass is essential
  for rejection of background. Based on these
  requirements, we formulated the design concepts for the 
calorimeter on the Monte-Carlo level and then built prototype detectors 
that were tested in the beam \cite{ashot2004}. Finally, we were able to
construct the PrimEx detector whose design and performance are
described below.
\section{Design of the Calorimeter}
 To optimize the performance and cost of the detector,
 we have chosen a hybrid calorimeter design combining novel $PbWO_4$
 scintillating crystals and conventional Cherenkov lead glass detectors.
 The central part of the calorimeter is built of 1152 $PbWO_4$ 
   crystals surrounded with 576 lead glass blocks. The transverse dimensions of
 the calorimeter ($116 x116 cm^2$) are sufficient for a large geometrical
 acceptance  for the $\pi^0$ mesons emitted at zero angle from the
production target located 7.3 m upstream of the detector ($\sim$70\%).
 Since the area 
 around the photon beam is under a high rate of irradiation, the crystals used should 
 be radiation hard. The $PbWO_4$ crystals fully meet the latter requirement.
 
 In the past 10-15 years, the $PbWO_4$ crystal development resulted in improved
 light output that provides a good energy resolution. Furthermore, due to very 
 strict requirements by LHC calorimetry groups \cite{lhc}, high quality 
 and radiation hard $PbWO_4$ crystals became commercially available from 
 the two basic manufacturers:
 BTCP, Russia \cite{btcp} and  Shanghai Institute of Ceramics (SIC), China.
 For the HYCAL calorimeter we have recieved
 1250 crystals from SIC, China.
 The dimensions of the individual crystals are
 20.5x20.5x180mm$^3$. To improve light collection, the crystals 
 were wrapped in $100\mathrm{\mu m}$ VM2000 reflective material \cite{ashot2004} and  
 then, for light isolation between the neighbours, in $36\mathrm{\mu m}$ Tedlar.
 The crystals were viewed by Hamamatsu R4125HA photomultiplier tubes 
 coupled to them with optical 
 grease. As the $PbWO_4$ crystal light yield is highly temperature dependent 
 ($\sim$-2\%/$^oC$ at room temperature), temperature stabilization of 
 the calorimeter is mandatory.
 Throughout the experiment, the calorimeter was operated at $14\pm0.1\mathrm{^oC}$ 
 temperature, which was maintained by circulation of the cool liquid around 
 the outer body of the calorimeter assembly.
 
 The lead-glass modules of the calorimeter were provided by the IHEP (Russia) 
 group \cite{lg}.
 The individual modules with the size of 38.2x38.2x450mm$^3$ (SF-2 type) were
 wrapped by $25\mathrm{\mu m}$ aluminized mylar foil.
  The Cherenkov light produced in the lead 
 glass radiator was detected by the Russian-made FEU-84-3 photomultiplier tubes.
\begin{center}
\begin{figure}
\caption{Photograph of the HYCAL calorimeter during assembly: 
crystal and lead glass arrays are shown}
\label{fig:calpix}
\includegraphics[angle=270,width=0.6\textwidth]{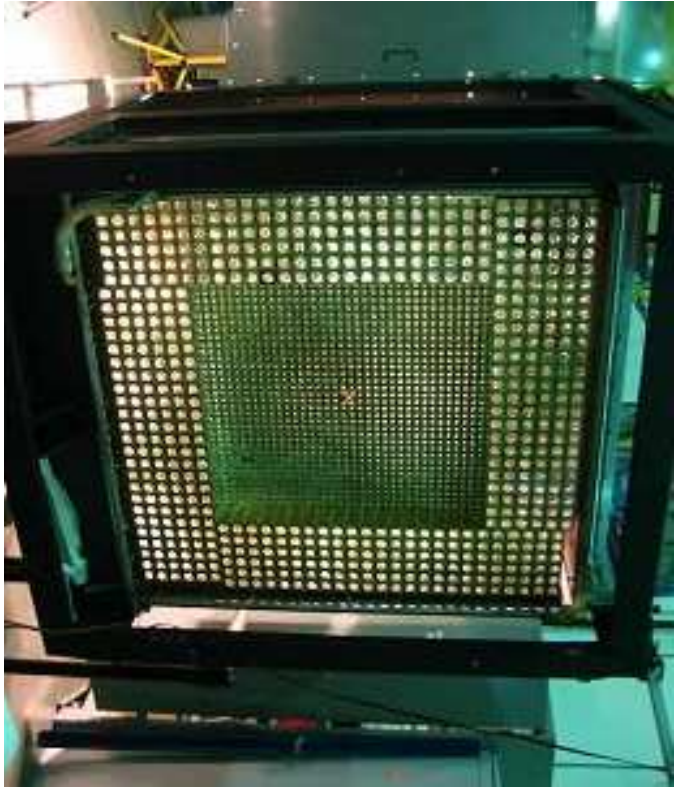}
\end{figure} 
\end{center}
  The photograph of the detector during the assembly at the TestLab of 
  the Jefferson laboratory is shown in Fig.~\ref{fig:calpix}.

 The PrimEx experiment utilized the CODA DAQ system. The digital information was read out
 from over 2200 channels of FASTBUS based ADC and TDC.
 An advantage of this system 
 is the ability to operate in fully 
 buffered mode. Namely, events are buffered in the digitization modules themselves
 allowing the modules to be "live" during the readout process. This significantly reduced the
 dead time of the DAQ, which is essential for the cross-section measurement.

\section{Calibration}

In order to maintain high performance of the calorimeter, a periodic energy 
calibration with tagged photon beam was required.
During the experiment,
HYCAL was calibrated by illuminating each module with a low-intensity 
tagged-photon beam.
The entire assembly with weight of about five tons was movable in the transporter frame in
both horizontal and vertical directions with a position accuracy of 
$\pm 0.7 mm$.  The calorimeter position was 
remotely controlled during the calibration.
\begin{center}
\begin{figure}
\includegraphics[angle=270,width= 1.0\textwidth]{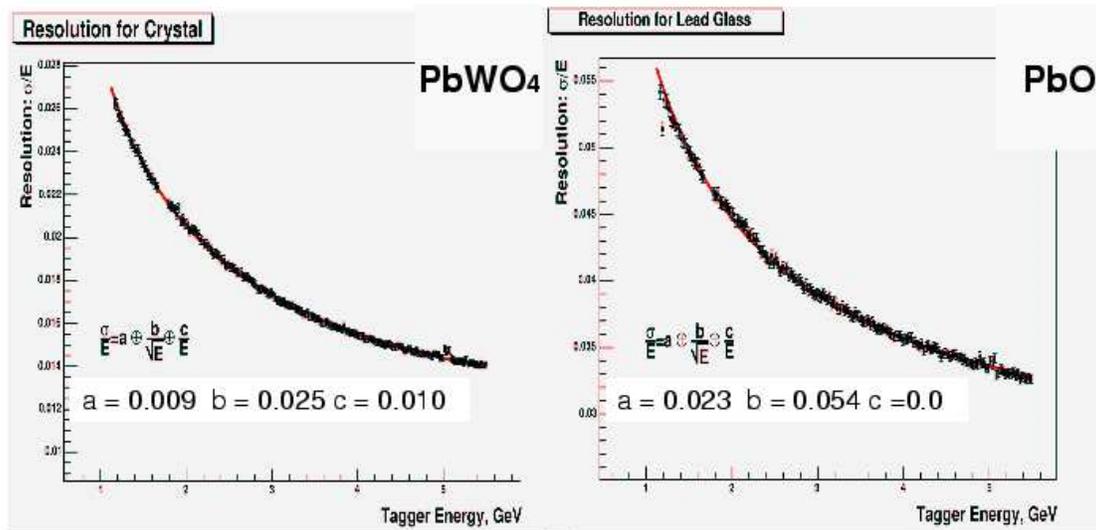}
\vspace{+0.5cm}
\caption{Energy resolution of the $PbWO_4$ crystals and the lead glass as 
function of the primary photon energy.}
\label{fig:res_cal}
\end{figure}
\end{center}
\begin{center}
\begin{figure}
\includegraphics[angle=270,width= 0.6\textwidth]{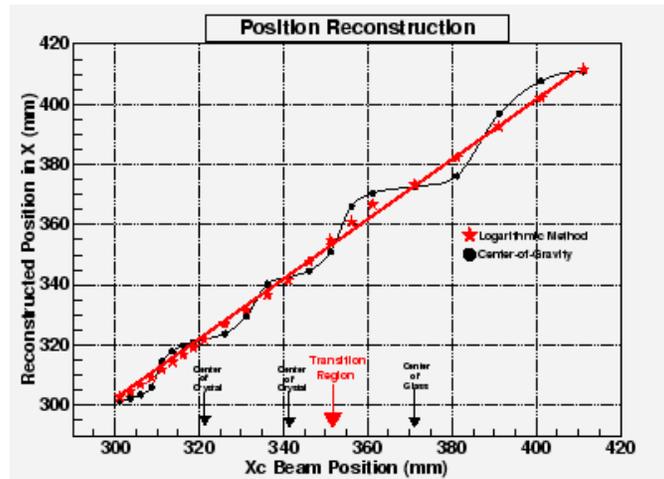}
\caption{Shower position reconstruction in the calorometer.}
\label{fig:posi}
\end{figure}
\end{center}
The tagged photon beam irradiated each module of the calorimeter 
using "a snake scan" during the
calibration runs. Relative gains of the modules 
and energy dependence of the detector resolution, as determined through
calibration, are shown in Fig.~\ref{fig:res_cal}
for the $PbWO_4$ and lead-glass parts of the detector.
The reconstruction of the $\gamma$ cluster position in the calorimeter 
is illustrated in Fig.~\ref{fig:posi}. The logarithmic method of coordinate 
reconstruction with fitted parameters was chosen.
The final fine-tuning of the modules' gains was made using photons 
from decays of neutral pions observed in the experiment.

In between the beam calibrations the gain stability of each channel was
monitored by a specially designed gain light monitoring system (LMS) \cite{prim_lms}.
The LMS consists of the following main parts: a light source,
a light mixing box, a light distribution system, and reference detectors with a DAQ
system.  The light source comprises an assembly of NICHIA super bright blue LEDs 
(peak wavelength 470 nm). Light mixing is done using a six inch diameter 
integrating sphere (ORIEL) that provided 2000 distribution channels. 
The pulsed light was distributed
to individual calorimeter modules via a bundle of plastic optical fibers. Each fiber 
is attached to the front face of a detector module. As a reference 
detector, a Hamamatsu PIN photodiode (S3399, $3\mathrm{mm}$ diameter) with an AMPTEC 
low noise charge amplifier was used. Light from the light monitoring system was 
periodically injected into the detector modules between the data
taking files ($\sim$1 hour period).
\section{Physics Run performance}
\begin{center}
\begin{figure}
\includegraphics[angle=0,width= 0.6\textwidth]{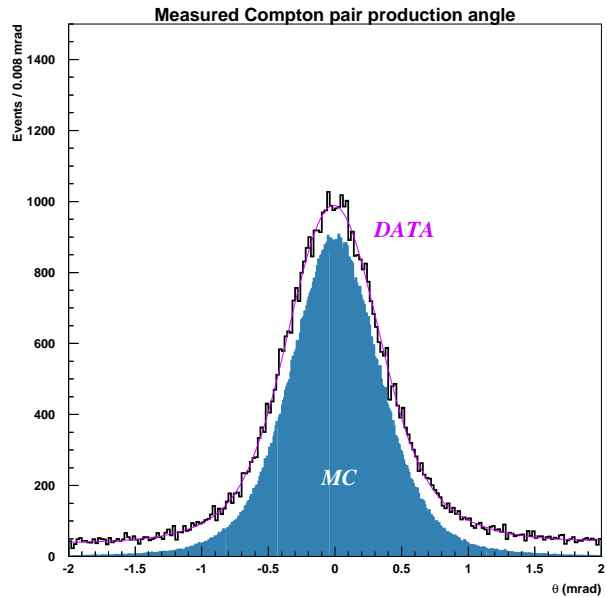}
\caption{Measured angular distribution of Compton-pair clusters compared with the 
Monte Carlo simulations: $\sigma_{theta} = 0.34 mrad$.}
\label{fig:car}
\end{figure}
\end{center}
The experiment took physics data in the fall of 2004. Several exposures with 
 beryllium, carbon, tin and lead targets were taken. The performance of the calorimeter 
 was studied using several observed physics processes: quasielastic and inelastic 
 photoproduction of neutral pions, Compton scattering and pair production.

\begin{center}
\begin{figure}
\includegraphics[angle=0,width= 0.8\textwidth]{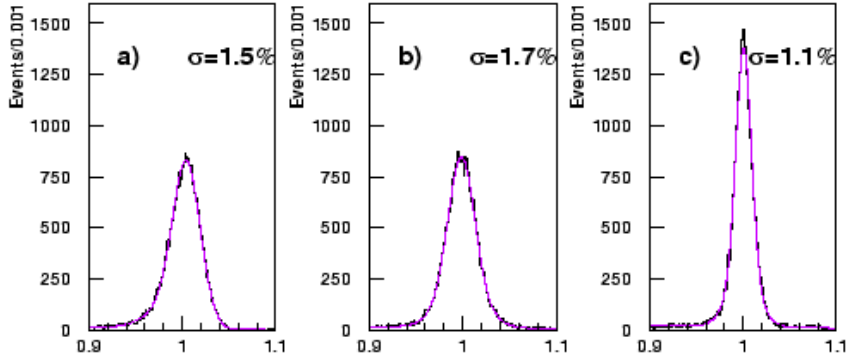}
\caption{Ratio of sum of measured Compton pair energy
and tagged $\gamma$ energy at 5.2 $\pm$ 0.3 GeV. a - using 
reconstructed cluster
energies; b - using cluster coordinates
and Compton kinematical relations to calculate Compton 
pair energy; c - after applying the transverse momentum constraint.}
\label{fig:p15}
\end{figure}
\end{center}
\begin{center}
\begin{figure}
\includegraphics[angle=0,width=0.8\textwidth]{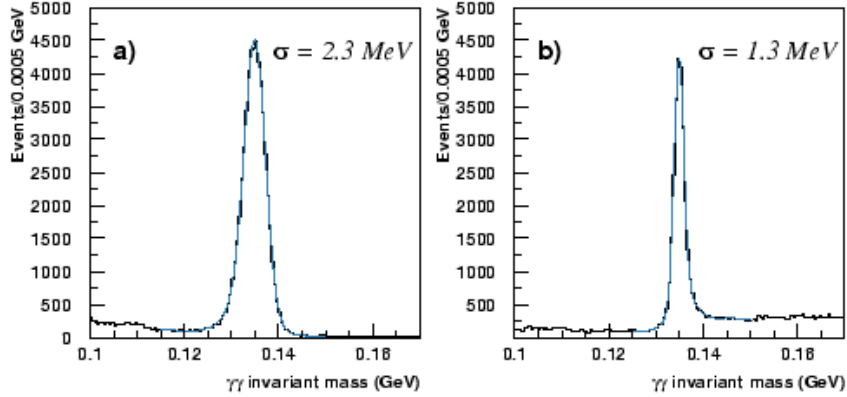}
\caption{$\pi^0$ mass resolution for the $PbWO_4$ array. 
Left: after calibration 
on the nominal $\pi^0$ mass; Right: 
after correction on the tagged photon energy:
the remaining width is mostly due to coordinate resolution.}
\label{fig:p16}
\end{figure} 
\end{center}
\begin{center}
\begin{figure}
\includegraphics[angle=0,width=0.8\textwidth]{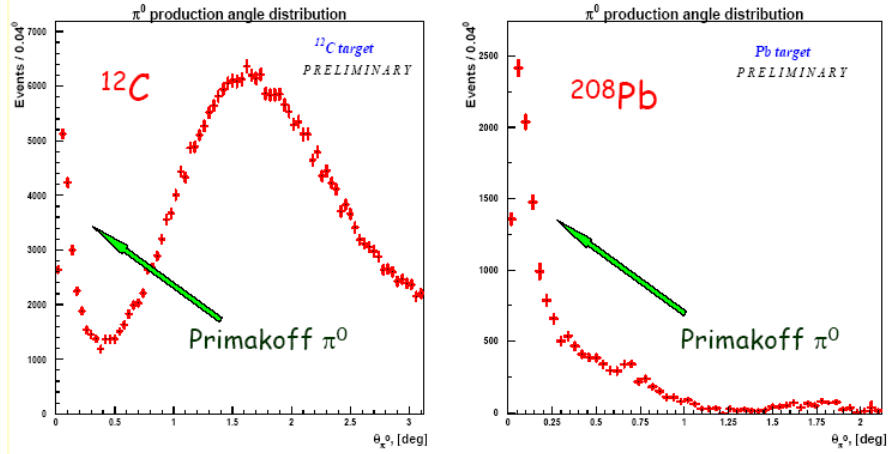}
\caption{$\pi^0$ angular distribution for production on Carbon and Lead. 
The Primakoff production peak is clearly visible at small angles.}
\label{fig:p19}
\end{figure} 
\end{center}
First, we consider Compton scattering $ \gamma + e \to \gamma + e $, which 
  provides four constraints of the 
  reaction kinematics and allows the study of energy and space resolutions of the 
  detector, albeit in a limited range of small angles for the central $PbWO_4$ 
  region of the calorimeter.
 Plotted in Fig.~\ref{fig:p15} is the ratio between 
  the total energy as measured by the calorimeter and 
  incident energy of the tagged photon ($\sim$5.2 GeV). 
  The width of the distribution is $\simeq1.5$\% in (a) where complete 
information from the calibration is used, and $\simeq1.7$\%
in (b) where the energy of the Compton pair is estimated only from the
resonstructed shower coordinates using the kinematics of Compton
scattering. The latter demonstrates the importance of good spatial
resolution. As soon as total transverse momentum of the two showers
is constrained to zero ($\overrightarrow{P_t} = 0$), the width of the
distribution reduces to mere 1.1\%, see Fig.~\ref{fig:p15}c.
As the Compton pairs are emitted at zero angle to the incident beam,
 the width of the distribution of the observed angle $\theta$ is a direct measure of 
 the angular 
  resolution of the detector. From the data of Fig.~\ref{fig:car} we estimate 
  the experimental resolution on $\theta$ as $\sigma_{\theta} = 0.34$ mrad, 
  which agrees with the simulated value and is sufficient for measuring 
  the Primakoff reaction.   

  Next, we estimate the detector resolution on the $\pi^0$ mass. 
  Fig.~\ref{fig:p16} shows the $\gamma-\gamma$ invariant mass distribution 
  for the central ($PbWO_4$) region of the calorimeter. The left-hand panel shows the 
  distribution upon calibration on the nominal mass of the 
  neutral pion (observed width of 2.3 $MeV/c^2$). In the right-hand panel,
  correction of the primary tagged photon energy has been applied. 
  The observed width of 1.3 $Mev/c^2$ in the latter case reflects the coordinate 
  resolution of the calorimeter. The constraint on the primary energy 
  significantly improves the mass resolution. In addition, we see a 
  high mass tail of the peak that is due to inelastic  $\pi^0$ production. 
  These pions have lower energy, and the beam-energy constraint shifts 
  their corrected masses towards higher values. 
  
  The gaussian width of the $\pi^0$ peak increases from 1.3 MeV/$c^2$ for the
$PbWO_4$ area of the calorimeter to 2.6 MeV/$c^2$ for the border area
(on the $PbWO_4$ side of it), and to 4.5 MeV/$c^2$ for the middle of the 
lead-glass area.
   
  The success of the experiment largely depends on accurate reconstruction
of $\pi^0$ mesons emitted at small angles. Shown in Fig.~\ref{fig:p19} 
are the preliminary data on $\pi^0$ angular distributions for quasielastic
production on carbon and lead nuclei. The Primakoff production peak is 
clearly seen for both data sets. The latter demonstrates good performance
of the calorimeter, which should be adequate for a precision 
measurement of the $\pi^0$ lifetime.
\section{Conclusion}
 The high precision hybrid electromagnetic calorimeter (HYCAL) has been designed, 
 constructed, and run in the experiment aimed at measuring the lifetime of the 
 neutral pion via the Primakoff effect. Measured characteristics of the detector 
 have been shown to match the design parameters. The energy resolution of the calorimeter is 
 about 1.5\% for a photon energy of 5 $GeV$. Angular resolution of 0.34 mrad 
 for Compton production is acheived. The $\pi^0$ mass resolution (upon constraining 
 the beam energy) is
 $\sigma = 1.3 MeV/c^2$ for the central PWO part of the calorimeter.
 High-quality 
 data sets for extracting the $\pi^0$ lifetime have been collected. The first results 
 are expected by the end of this summer. 
 The authors thank all PrimEx collaborators for excellent
 design, construction, running, and data analysis of the experiment. 
 Stable operation of the CEBAF accelerator is gratefully  acknowledged. 
 The work was supported by the US NSF grant (PHY-0079840) and the RFBR grant 
 (04-02-17466).
\doingARLO[\bibliographystyle{aipproc}]
          {\ifthenelse{\equal{\AIPcitestyleselect}{num}}
             {\bibliographystyle{arlonum}}
             {\bibliographystyle{arlobib}}
          }
\bibliography{sample}

\end{document}